\documentclass[useAMS,usenatbib]{mn2e}

\title[Constraints on fall-back disks]{Constraints on fall-back disks in
radio pulsars and anomalous X-ray pulsars}
\author[P. B. Jones]{P. B. Jones\thanks{E-mail:
p.jones1@physics.ox.ac.uk}\\
University of Oxford, Department of Physics, Denys Wilkinson building,
Keble Road, Oxford OX1 3RH, England\\}
\begin{document}

\date{ }

\pagerange{\pageref{}--\pageref{}} \pubyear{}

\maketitle

\label{firstpage}

\begin{abstract}
Calculations have been made of fall-back disk heating by the pulsar
wind as distinct from the soft X-rays emitted by the neutron-star
surface.  The relation between these heating rates and measured
near-infrared fluxes in the K and
K$_{\rm s}$-bands places severe constraints on the inner radii of
any fall-back disks that may be present in radio pulsars and in some 
anomalous X-ray pulsars.  The lower limits found are so large that
the disks concerned can have no significant effect on pulsar spin
down.
\end{abstract}


\section{Introduction}

Dust and debris disks are a commonplace phenomenon in astrophysics,
being a feature of the major planets and of pre-main-sequence stars
(see, for example, Beckwith et al 1990).  The properties of dust disks
or of viscous ionized disks formed as a consequence of fall-back in
supernovae have been a matter of continuing interest in neutron-star
physics.  But until recently, there has been no observational evidence
that they exist. For a bibliography, we refer to
Perna, Hernquist \& Narayan
(2000), Menou, Perna \& Hernquist (2001), Blackman \& Perna (2004),
Ek\c{s}i \& Alpar (2005), Ek\c{s}i, Hernquist \& Narayan (2005) and to the 
many earlier papers on the subject
cited by these authors.

Further interest has been stimulated by flux measurements on the
anomalous X-ray pulsar 4U 0142+61 in the 4.5 and
8.0 $\mu$m infrared bands (Wang, Chakrabarty \& Kaplan 2006). The energy
spectrum derived from these measurements, and from earlier K$_{\rm s}$-band
and optical fluxes, is consistent with blackbody emission from a dust
disk with an inner radius surface temperature of 1200 K, superimposed on
a power-law spectrum.  The optical emission, which is pulsed at the
neutron-star rotation frequency, consists almost entirely of the
power-law component, which is assumed to be of magnetospheric origin.

There are also flux limits at 4.5, 8.0 and 24 $\mu$m in the
mid-infrared for a limited
number of isolated neutron stars (Wang, Kaspi \& Higdon 2007) and for
some supernova remnants at 4.5 and 8.0 $\mu$m
(Wang, Kaplan \& Chakrabarty 2007).  But in the K and K$_{\rm s}$-bands,
a considerable number of magnitudes have been measured.
These have been summarized by Mignani et al (2007) and include
radio pulsars, anomalous X-ray pulsars (AXP), and soft-gamma
repeaters (SGR). The purpose of the
present paper is to see how these measurements constrain possible
fall-back disk parameters.

Fall-back disk formation is a detail of the supernova event and no attempt
is made to describe it here.  Initially, the disk must have been internally
ionized and hence viscous, with mass and angular momentum transfer rates
intrinsic to its formation.  We assume that at some later time, a 
transition
to a passive disk occurs through the the thermal ionization instability
(see Menou et al 2001) in which free electron recombination produces a 
sudden
decrease in opacity which propagates rapidly inward.  The disk formed has
little internal ionization and viscosity, and is gaseous. Its
luminosity is then principally a result of interaction with the pulsar wind
or with
blackbody radiation from the neutron star surface.  The distribution of 
mass
between dust grains and gaseous molecules depends on the various 
sublimation
temperatures concerned.  Our assumption is that, at the time of 
observation,
the inner radius of the disk lies outside the light-cylinder radius 
$R_{LC}$
of the neutron star.  An important consequence is that we can treat the 
disk
as a partial termination of the pulsar wind rather than as a boundary 
condition
to be satisfied by the solution for the magnetospheric fields.  

A previous paper (Jones 2007) on the interaction of a thin passive disk 
with the
pulsar wind described the ablation processes in some detail and calculated
the alignment and precession torques acting on the disk.  Section 2 of the
present paper extends the discussion of disk luminosity.  The K and
K$_{\rm s}$-band magnitudes listed by Mignani et al for sets of radio 
pulsars
and
AXP or SGR are used, in Section 3, to constrain the inner radii and
temperatures of passive fall-back disks that might be present in these
objects.  Where possible, the predicted mid-infrared fluxes from these
objects are compared with the flux limits found by Wang, Kaspi \& Higdon.

\section[]{Interaction with the pulsar wind}

The Deutsch vacuum solution for the electromagnetic fields ${\bf E}$ and
${\bf B}$ of a rotating neutron star (see Michel \& Li 1999, Ek\c{s}i \&
Alpar 2005) makes it possible to write down explicit expressions for the
spherical polar components of the momentum density ${\bf p}$,
\begin{eqnarray}
{\bf p}=\frac{1}{4\pi c}{\bf D}\times{\bf B},
\end{eqnarray}
in the inertial frame at radii $r > R_{LC}$, the z-axis being parallel
with the neutron-star spin angular velocity ${\bf \Omega}$.
The momentum density in a
physical pulsar wind, in which both fields and relativistic particles are
present, may differ from this in detail, but the time-averaged long-range
part of the azimuthal component $\langle{\bf p}_{\phi}\rangle$ must be
finite because it is directly related to the
neutron-star spin-down torque (see Section 2.1 of Jones 2007). The
composition of the wind does not
significantly affect its interaction with the disk and it is
therefore correct to use this quantity derived from the vacuum solution
in calculations of ablation and, in particular, of alignment
and precession torques.

The vacuum-solution time-averaged
polar component is $\langle{\bf p}_{\theta}\rangle = 0$ for all angles,
but its uni-directional time average, taken in the direction of either
increasing or decreasing $\theta$, can contribute to heating of the disk
surface though not to a torque.  With the addition of this component,
the power input per unit
area, to the one side of a thin disk that is viewed by an observer,
can be written down and equated with the
blackbody radiation emitted at the local temperature $T(r)$,
\begin{eqnarray}
\sigma T^{4}(r) = \frac{3\mathcal{L}_{w}R_{LC}}
{8\pi^{2}r^{3}}\left(\sin\beta + \frac{2}{3}\cot\xi\right),
\end{eqnarray}
where $\mathcal{L}_{w}$ is the pulsar wind luminosity, $\beta$ is the disk
tilt angle and $\xi$ is the angle between the neutron-star magnetic and
spin axes. The first term in equation (2) is derived from the long-range
part of $\langle{\bf p}_{\phi}\rangle$ by means of equation (14) of
Jones (2007).  The second term, in the approximation of small
$\beta$, is from the long-range part of the
uni-directional time average of ${\bf p}_{\theta}$ evaluated in the
vicinity of $\theta = \pi/2$.  The status of the vacuum solution form
of this term is
much less certain.  Apart from an apparent unphysical singularity at
$\xi =0$,
which is a consequence of the vacuum field solution and our use of
$\mathcal{L}_{w} \propto \sin^{2}\xi$ as a working parameter, the term is
linearly dependent on the radial component of the electric displacement
${\bf D}$ which is undetermined because the total charge of the star and
the physical, as opposed to vacuum, magnetosphere is an unknown quantity,
as has been emphasized by Michel \& Li (1999).

The values that might be found for the tilt angle $\beta$ are, of course,
unknown, but there is a case for assuming that its value at formation is
non-zero as a consequence of the fact that type II supernovae are generally
asymmetric (see Goldreich, Lai \& Sahrling 1997). This suggests that the
angular momenta of the neutron star, the fall-back disk and the ejected
matter may not all be precisely parallel.

The uncertainties in $\beta$ and in the polar term are so considerable that
we shall replace the vacuum-field quantity
$\sin\beta + 2/3\cot\xi$ in equation (2) by a free parameter
$\zeta$ and adopt a conservative range of values
$0.03 \leq \zeta \leq 0.3$.  Neglect of the wind albedo is a further
uncertainty in equation (2), but not one that is serious because
relativistic
particles either in the wind or accelerated in the diffuse surface of the
disk as described previously (Jones 2007) are attenuated only at depths of
$10^{2}-10^{3}$ g cm$^{-2}$ within the disk. (We assume that the mean
disk density $\bar{\Sigma}$ exceeds $10^{3}$ g cm$^{-2}$.)
For the same reason, it is not anticipated
that infrared or optical emission from  an undistorted disk would have an
observable component pulsed at the neutron-star rotation period.

We anticipate that, even for young neutron stars, almost all of the 
potential
disk ablation will have occurred by the time of observation.
Interaction with the pulsar wind increases the inner radius $r_{i}$
which is given by
\begin{eqnarray}
r_{i}^{3}(t) = \frac{3cIm_{H}\mathcal{N}_{\epsilon}\zeta
(\Omega_{0} - \Omega(t))}{4\pi^{2}\bar{\Sigma}},
\end{eqnarray}
which is valid provided the solution $r_{i}(t)$ exceeds the real $r_{i}$
at formation.
Here $\Omega_{0}$ is the neutron-star spin angular velocity at formation,
$I$ is the neutron-star moment of inertia and $m_{H}$ the mass of the 
hydrogen
atom. The remaining parameter $\mathcal{N}_{\epsilon}$ is the number of 
baryons
(mostly neutrons) removed from the disk per unit incident wind energy
(see Jones 2007; equations 17 and 18).  The mean mass density 
$\bar{\Sigma}$
is the initial value at formation.  Owing to the number of unknown 
parameters
it contains, equation (3) does not provide a useful constraint on disk 
inner
radii derived from equation (2). But it shows that, in most cases, the 
existing $r_{i}$ must be near its $\Omega = 0$ limit.  Because equation (3) 
is based on
the interaction of the disk with the pulsar wind, it ceases to be valid
when the light-cylinder radius exceeds $r_{i}$. But for the radio pulsars,
this occurs only for $\Omega$ so small as to be of no interest.

\section[]{Disk parameters}

Recent observations in the near-infrared have produced H, K and
K$_{\rm s}$-band magnitudes for many isolated neutron stars, including
both radio pulsars and AXP. A useful listing of these has been published
by Mignani et al (2007) who also include estimates of the distances $d$.
We have used the Spitzer Science Centre magnitude to flux-density
converter to obtain the observed energy fluxes per unit frequency
interval $f_{\nu}$.  Equation (2) provides the disk surface temperature
for each object as a function of radius and it is then straightforward,
by integration and
interpolation and with a number of moderate assumptions, to find the
disk inner radius that gives $f_{\nu}$.  In particular, we assume that
the angle
$\theta_{\perp}$ between the disk normal and the line of sight is fixed
and given by $\cos\theta_{\perp} = 0.5$.  No attempt has been made to
correct for the small K-band interstellar
extinction.  The values found for $r_{i}$ are quite
insensitive to the outer radius $r_{o}$ because,
for the temperatures found here, the outer
part of the disk contributes little to the K-band flux.  Therefore, we
have adopted a fixed $r_{o} = 3.4r_{i}$, equal to the best-fit ratio found
by Wang et al in the case of 4U 0142+61.  With these assumptions, values of
$r_{i}$ and the inner radius temperature $T_{i}$ have been computed and are
given in the right-hand six columns of Table 1 for $\zeta = 0.03, 0.1, 
0.3$.

The disks of Table 1 need not be massive: for $\zeta = 0.03$, the masses
are in the interval $10^{-7}$ - $10^{-5}$ M$_{\odot}$, assuming a density
$\bar{\Sigma} = 10^{3}$ g cm$^{-2}$, but their
angular momenta can be of the same order as that of the neutron star.
Certainly for the lower temperatures given in Table 1, the disks must
consist almost entirely of dust grains and only a small gaseous
component.  But even a very low
concentration of free electrons and ions is sufficient to stop the
pulsar wind (see Jones 2007) and there can be no doubt that such
concentrations must be present from a variety of sources.

For consistency with our Section 1 definition of a fall-back disk, we
require that a satisfactory solution has $r_{i} > R_{LC}$.  The first
six rows in the Table refer to radio pulsars, and they satisfy this
condition
by large margins. But it is interesting to see that, in many cases, the
AXP and SGR do not.  In particular, there is no solution for 4U 0142+61,
the object for which evidence of a disk is most compelling.  This is not a
cause for concern because equation (2) contains only the pulsar wind
contribution to disk heating.  For the AXP and SGR, the soft X-ray
luminosity $\mathcal{L}_{bol}$, possibly of thermal origin, exceeds
$\mathcal{L}_{w}$ by several orders of magnitude so that its power input
can be the more important if the disk is not of negligible thickness but
subtends a finite solid angle at
the neutron star, as in the model adopted by Wang et al (2006).

It is reasonable to assume that a disk, immediately
after the ionization instability transition, would be plane and thin.
If there is a gaseous component, it would have a diffuse surface of
approximate depth
\begin{eqnarray}
h = \sqrt{\frac{k_{B}T}{m_{A,Z}\Omega^{2}_{K}}},
\end{eqnarray}
where $m_{A,Z}$ is an atomic (or molecular) mass and $\Omega _{K}$
is the Kepler angular frequency. For the 4U 0142+61 best-fit inner
radius $r_{i} = 2.0\times 10^{11}$ cm found by Wang et al, the very
small Kepler frequency, $1.5\times 10^{-4}$ rad s$^{-1}$ leads to a
fairly large depth, $h \approx 2\times  10^{9}$ cm, for atomic hydrogen
at $T = 10^{3}$ K.  The 4U 0142+61 luminosity is
$\mathcal{L}_{bol} \sim 10^{35}- 10^{36}$, so that although the disk is
thin, the X-ray power input, $(1-\eta_{d})h\mathcal{L}_{bol}/r_{i}$
exceeds that given by equation (2) even for the large value of the X-ray
albedo $\eta_{d}$ adopted by Wang et al (2006).  The solid angle
subtended by the disk
may also be increased by departures from the plane assumption
arising from $r$-dependent alignment
torques (Jones 2007) or disk instability
(Petterson 1977; Pringle 1996).  The X-ray contribution to disk
heating is therefore important for the AXP and SGR, but not for the
radio pulsars which have $\mathcal{L}_{bol} \ll \mathcal{L}_{w}$.

\begin{table*}
\centering
\begin{minipage}{170mm}
\caption{Disk inner radii $r_{i}$ that are consistent with measured
K and K$_{\rm s}$-band magnitudes have been calculated, under the 
assumptions
made in Section 3, for a number of radio pulsars and AXP or SGR. Inner
radii temperatures $T_{i}$ are also given.  For J0142+61, J1810-197,
IE 2259+586,
and SGR 1806-20, the values of the wind luminosity $\mathcal{L}_{w}$
have been obtained from the periods $P$ and spin-down rates given,
respectively, by
Gavril \& Kaspi (2002), Ibrahim et al (2004), Woods et al (2004) and
Mereghetti et al (2005). For the remaining neutron stars, the periods and
$\mathcal{L}_{w}$ are from the Australia Telescope National Facility
Pulsar Catalogue (Manchester et al 2005).  The distances $d$ are those
listed by Mignani et al (2007).  The energy flux densities
$f_{\nu}$ have been obtained from the observed K or K$_{\rm s}$-band
magnitudes using the Spitzer Science Centre magnitude to flux-density
converter.  The right-hand six columns give $r_{i}$ and $T_{i}$ for
$\zeta = 0.03, 0.1, 0.3$.  Blank spaces indicate the absence of a solution
with $r_{i} > R_{LC}$. }
\begin{tabular}{@{}lrrrlrrrrrr@{}}
\hline
neutron star & $P$ & $\mathcal{L}_{w}$ & $f_{\nu}$ & d & $r_{i}$ & $T_{i}$ 
&
 $r_{i}$ & $T_{i}$ & $r_{i}$ & $T_{i}$  \\
    & s & erg s$^{-1}$ & erg cm$^{-2}$ & kpc & cm & K &
	cm & K & cm & K \\
	& & & s$^{-1}$ Hz$^{-1}$ & & \multicolumn{2}{c}{$\zeta=0.03$} &
	\multicolumn{2}{c}{$\zeta=0.1$} & \multicolumn{2}{c}{$\zeta=0.3$}  \\
 
\hline

B0531+21 & 0.033 & $4.6\times 10^{38}$ & $2.0\times 10^{-26}$ & 1.73 &
 $4.6\times 10^{11}$ & 1970 & $1.0\times 10^{12}$ & 1440 &
 $2.0\times 10^{12}$ & 1180 \\

B0633+17 & 0.237 & $3.2\times 10^{34}$ & $2.9\times 10^{-30}$ & 0.16 &
 $2.5\times 10^{11}$ & 470 & $4.0\times 10^{11}$ &
  440 & $6.2\times 10^{11}$ & 420  \\
  
B0656+14 & 0.385 & $3.8\times 10^{34}$ & $6.1\times 10^{-30}$ & 0.29 &
 $2.6\times 10^{11}$ & 540 & $4.2\times 10^{11}$ & 500 &
 $6.6\times 10^{11}$ & 470  \\
 
B0833-45 & 0.089 & $6.9\times 10^{36}$ & $2.0\times 10^{-29}$ & 0.29 &
 $1.0\times 10^{12}$ & 480 & $1.7\times 10^{12}$ & 450 &
 $2.6\times 10^{12}$ & 420  \\
 
B1509-58 & 0.151 & $1.8\times 10^{37}$ & $1.2\times 10^{-28}$ & 4.2 &
 $6.2\times 10^{11}$ & 1020 & $1.1\times 10^{12}$ & 880 &
 $1.9\times 10^{12}$ & 780  \\

J1119-6127 & 0.408 & $2.3\times 10^{36}$ & $<1.1\times 10^{-29}$ &
 6.0 & $>5.5\times 10^{11}$ & $<860 $ & $>9.7\times 10^{11}$ & $<760$ &
 $>1.6\times 10^{12}$ & $<690$  \\
 
\vspace{5mm} \\

J0142+61 & 8.69 & $1.2\times 10^{32}$ & $6.1\times 10^{-29}$ & $\ge 5$ &
 - & - & - & - & - & -  \\
 
J1048-5937 & 6.45 & $5.6\times 10^{33}$ & $2.0\times 10^{-29}$ & 3 &
 $1.3\times 10^{11}$ & 1110 & $2.4\times 10^{11}$ & 950 &
 $4.1\times 10^{11}$ & 840  \\
 
J1708-4009 & 11.00 & $5.8\times 10^{32}$ & $3.2\times 10^{-28}$ & 5 &
- & - & - & - & - & -  \\
 
J1810-197 & 5.54 & $2.7\times 10^{33}$ & $3.2\times 10^{-29}$ & 4 &
 $5.0\times 10^{10}$ & 1840 & $1.1\times 10^{11}$ & 1380 &
 $2.0\times 10^{11}$ & 1140  \\
 
IE 2259+586 & 6.98 & $5.6\times 10^{31}$ & $1.4\times 10^{-29}$ & 3.0 &
- & - & - & - & $4.2\times 10^{10}$ & 1510  \\
 
SGR 1806-20 & 7.56 & $5.0\times 10^{34}$ & $6.1\times 10^{-29}$ &
 15 & - & - & $1.7\times 10^{11}$ & 2180 & $4.0\times 10^{11}$ & 1570  \\

 \hline
 \end{tabular}
 \end{minipage}
 \end{table*}

\section[]{Comparison with observation}

The electromagnetic spectra of a limited number of radio pulsars are
known over wide frequency intervals.  In the case of B0656+14, as an
example, apart from the lack of measurements between $10^{10}$ and
$10^{14}$ Hz, the shape of the spectrum is of blackbody
emission from the neutron-star surface superimposed on a 
background which decreases monotonically with frequency and is
possibly of
some common magnetospheric origin (see Koptsevich et al 2001; Fig. 6)
The K-band flux listed in Table 1 is consistent with this spectrum. 
It is also known that the optical emission is pulsed at the rotation
frequency.  For example, Shearer et al (1997), observing B-band
emission, found a $1\sigma$ upper limit of only
$8\times 10^{-31}$ erg s$^{-1}$ cm$^{-2}$ Hz$^{-1}$
on the unpulsed component of the flux.  With regard to the AXP listed in
Table 1, the optical emission from 4U 0142+61 has a
pulsed fraction of 0.3, and the optical and X-ray pulse profiles
are in phase and of similar shape
(Kern \& Martin 2002, Dhillon et al 2005).

Unambiguous evidence for the presence of a disk would be an unpulsed
blackbody excess which, for the temperatures given in Table 1, would
probably be best observed in the 4.5 and 8.0 $\mu$m bands.  With the
possible exception of 4U 0142+61, existing spectra provide no such
evidence.

Even for the smallest value of $\zeta$ assumed in Table 1, the radio
pulsar disk
inner radii that would be compatible with the K-band fluxes are all
larger than the best-fit $r_{i} = 2.0\times 10^{11}$ cm found by
Wang et al (2006) in the case of 4U 0142+61, although they are of
a similar order of magnitude.  It might be argued that
the correct value of $\zeta$ is much
smaller than $0.03$, perhaps by one or more orders of magnitude.
For this to be so, the neutron
star spin ${\bf \Omega}$ would have to be almost exactly normal
to the plane of the disk.
Also, the star and magnetosphere must be electrically
neutral to the extent that the uni-directional
time-average of ${\bf p_{\theta}}$, evaluated at $\theta = \pi/2$,
is two or more orders of magnitude smaller than the vacuum solution
value.  Instead, we suggest that the inner radii given for the radio
pulsars in Table 1 are broadly correct as lower limits.  

Heating of a disk with inner radius $r_{i} > R_{LC}$ by the pulsar wind
is capable of producing the observed $K_{\rm s}$-band fluxes in only two
of the AXP listed in Table 1.  But these objects are differ significantly
from radio pulsars in having very high soft X-ray luminosities,
$\mathcal{L}_{bol} \gg \mathcal{L}_{w}$, which give most of the disk
heating, as in the model adopted by Wang et al (2006).  The two
exceptional AXP, J1048-5937 and J1810-197,  are interesting in that
$3\sigma$ upper limits for their
mid-infrared fluxes at 4.5, 8.0 and 24 $\mu$m have been published
recently by Wang, Kaspi \& Higdon (2007).  These fluxes can also be
calculated from the $r_{i}$ and $T_{i}$ values given in the Table. 
For J1810-197, all the predicted fluxes are below the the $3\sigma$
upper limits.  However, the J1048-5937 4.5 $\mu$m $3\sigma$ upper limit
of $8\times 10^{-29}$ erg cm$^{-2}$ s$^{-1}$ Hz$^{-1}$ is exceeded by
the predicted flux of
$1.2\times 10^{-28}$ erg cm$^{-2}$ s$^{-1}$ Hz$^{-1}$ for
$\zeta = 0.03$ which increases to
$3.1\times 10^{-28}$ erg cm$^{-2}$ s$^{-1}$ Hz$^{-1}$ at $\zeta =0.3$.
A very small value, $\zeta = 0.012$ is required to fit both the
measured K$_{\rm s}$-band flux and the published $3\sigma$ upper limit.
In this case, the disk is at a higher temperature $T_{i} = 1300$ K and
has a smaller radius $r_{i} = 7.9\times 10^{10}$ cm.  But an alternative
possibility is that the source of both the K$_{\rm s}$-band and 4.5 $\mu$m
fluxes is magnetospheric rather than a heated disk, which could be
confirmed by a search for a pulsed component.

\section[]{Conclusions}

The present paper considers heating of a thin passive disk by non-radial
components of the pulsar wind and computes inner radii and temperatures
that are compatible with the observed K and K$_{\rm s}$-band fluxes for a
number of radio pulsars and AXP. In the case of the radio pulsars, the
inner radii are two or more orders of magnitude larger than the $R_{LC}$
and the disks concerned, if they exist, can have no significant effect
on pulsar spin-down. The conservative, and probably incorrect, assumption
made is that the observed K and K$_{\rm s}$ fluxes have no pulsed component
and are of thermal origin.  It is also assumed that there is no internal
disk viscosity and that heating by the radial flux of soft X-rays from the 
neutron-star surface is negligible except in the case of the AXP.

The possibility that some of the individual neutron stars listed in
Table 1 have disks has been investigated by previous authors.
K-band fluxes from disks with inner radii near the Alfv\'{e}n surface have 
been calculated in the case of B0656+14, and the four AXP J0142+61, 
J1048-5937, J1708-4009 and IE 2259+586, by Perna, Hernquist \& Narayan 
(2000). Also, the radio pulsars B0531+21 and B0833-45 were considered by 
Blackman \& Perna (2004).  The disk model differs from that of Section 2 in 
that the sources of heating were restricted to internal viscous dissipation 
and the soft X-ray spectrum of the neutron-star surface.  Even so, the 
predicted K-band fluxes for the four AXP were $f_{\nu} \sim 10^{-27}- 
10^{-26}$ erg cm$^{-2}$ s$^{-1}$ Hz$^{-1}$, considerably higher than the 
observed values listed in Table 1. In the case of the three radio pulsars, 
inclusion of heating by non-radial wind components is crucial and leads us 
to the conclusion that the proposed disks are not consistent with the 
observed K-band fluxes.  This is also true in the general case, treated by 
Ek\c{s}i \& Alpar (2005), of disks with inner radii near the light-cylinder 
radius. 
This conclusion should perhaps be qualified because we do not consider 
disks with $r_{i} < R_{LC}$.  Thus a disk satisfying this condition at 
formation, with $R_{LC}$ given by the initial angular frequency 
$\Omega_{0}$, would have its middle section removed by wind ablation 
leaving an outer section with the properties given in Table 1 and an inner 
section  completely within the present light-cylinder radius.  This inner 
sector would form part of the closed magnetosphere and would not intersect 
open magnetic flux lines or interact with the pulsar wind.  The properties 
of such remnant disks, if they exist, are not considered in this paper.

\bsp

\label{lastpage}

\end{document}